\documentclass[journal, 10 pt, a4paper]{IEEEtran}

\newtheorem{thm}{Theorem}
\newtheorem{cor}{Corollary}
\newtheorem{lem}{Lemma}

\usepackage[final]{graphicx}
\usepackage[reqno]{amsmath}
\usepackage{amssymb}
\usepackage{subfig}
\usepackage{epstopdf}
\usepackage{bm}

\begin{document}

\sloppy

\title{Topological Interference Management: Linear Cooperation is not useful for Wyner's Networks}
\author{\IEEEauthorblockN{Aly El Gamal\\}
 \IEEEauthorblockA{ECE Department, Purdue University\\ Email: \{elgamala\}@purdue.edu}}

\maketitle

\begin{abstract}
In this work, we study the value of cooperative transmission in wireless networks if no channel state information is available at the transmitters (no CSIT). Our focus is on large locally connected networks, where each transmitter is connected to the receiver that has the same index as well as $L$ succeeding receivers. The cases of $L=1$ and $L=2$ represent Wyner's asymmetric and symmetric network models, respectively. The considered rate criterion is the per user Degrees of Freedom (puDoF) as the number of transmitter-receiver pairs goes to infinity. For the case when $L=1$, it was shown in previous work that linear cooperation schemes do not increases the puDoF value, and that the optimal scheme relies on assigning each message to a single transmitter and using orthogonal access (TDMA). Here, we extend this conclusion to the case where $L=2$, by proving optimality of TDMA in this case as well. We conclude by discussing whether increasing the value of $L$ can create a value for linear cooperation schemes from a DoF perspective.

\end{abstract}

\section{Introduction}
%High-level overview of future wireless networks
It is expected that we witness a rapid increase in the size of wireless networks as well as the nature of their applications in the coming decade. With the deployment of self-driving cars, there are anticipated developments in vehicular communication networks to manage how vehicles communicate with each other (Vehicle-to-Vehicle or V2V networks) as well as how vehicles communicate with the infrastructure (Vehicle-to-Infrastructure or V2I networks). Wireless sensor networks are also planned to be used in a wide range of applications from military deployments to aid in emergency situations. Two distinctive features of networks serving these applications are their ad-hoc nature and the unusual short life-time of the network. These features make it increasingly difficult in practice to rely on the availability of accurate channel state information at the transmitters (CSIT) for the success of the used communication scheme. Further, even for standard cellular networks, these same features could arise because of increased deployments of heterogeneous networks, as well as the potential presence of deep fading conditions that could alter the network topology.

The focus of this work is on communication in single-hop wireless networks with no CSIT. The performance criterion we use is the degrees of freedom (DoF) or the pre-log factor of the sum capacity at high signal to noise ratio. DoF analyses are attractive for theoretical studies because they simplify the problem of capacity characterization and offer insights on the design of optimal coding schemes. However, it is important to note that DoF analyses are insensitive to variations in the strength of different channels in the network, and the conclusions we are drawing here may not extend to the case where channel strengths are different, as for example the Generalized Degrees of Freedom (GDoF) analysis in~\cite{Gesbert-weakcsit}.

%Previous work on TIM (not clear the impact on Sum DoF)
The term \emph{Topological Interference Management} (TIM) has been used in the literature to describe the problem of studying the DoF of interference networks with no CSIT. Special emphasis has been given to characterizing the symmetric DoF in this setting. By symmetry, we mean the DoF value that can be achieved simultaneously for all users. In~\cite{jafar-topological}, this problem was studied in a setting where channels remain constant, and a \emph{blind interference alignment} scheme was shown to achieve symmetric DoF gains that cannot be achieved by TDMA schemes. In~\cite{Naderi-ElGamal-Avestimehr} and~\cite{ElGamal-Naderi-Avestimehr}, this problem was studied in a setting where the channel is drawn randomly from a continuous distribution in each time slot (time-varying channel), and a class of \emph{retransmission-based} schemes was identified to be optimal in many scenarios and was shown to achieve symmetric DoF gains that cannot be achieved by TDMA schemes.

In this work, we study the sum DoF of large locally connected networks with no CSIT. The question we raise is whether cooperative transmission is useful in this setting. To the best of our knowledge, this is among the first few attempts to understand optimal sum rates in large networks with no CSIT. In~\cite{Gesbert-nocsit}, the DoF with no CSIT was studied under the assumption that a message could be delivered to its destined receiver through any of the transmitters connected to it. This assumption could reflect a \emph{flexible cell association} scenario in cellular downlink, where mobile terminal receivers can be associated with any base station transmitter connected to it. It was shown that the possibility of flexible cell association can be beneficial - from a sum DoF perspective - even with no CSIT, over a fixed cell association scenario where each receiver must be served by the transmitter carrying the same index. This was shown to be the case for a wide class of network topologies. In~\cite{ElGamal-spawc}, the DoF of large Wyner's asymmetric networks (channel model is introduced in~\cite{Wyner}) was characterized with no CSIT, and it was shown that TDMA is optimal, even if each message can be available at multiple transmitters and linear cooperation schemes are allowed. 

We start this work by discussing related work in Section~\ref{sec:related}, and then we state the system model in Section~\ref{sec:model}. In Section~\ref{sec:coop}, we extend the work of~\cite{ElGamal-spawc} to large networks where each transmitter is connected to the receiver with the same index as well as two succeeding receivers. Since we are only interested in studying large networks, the same results we obtain apply to a Wyner's symmetric network model, where each transmitter is connected to its own receiver as well as one preceding receiver and one succeeding receiver. Each message is allowed to be available at all transmitters and linear cooperation schemes are allowed. We show that even with these options, assigning each message to only one transmitter and employing a TDMA scheme is DoF-optimal. We then study in Section~\ref{sec:general} the problem for a general locally connected network where each transmitter is connected to the receiver with the same index as well as $L$ succeeding receivers, and $L > 2$. We provide concluding remarks in Section~\ref{sec:conclusion}.

\subsection{Related Work}\label{sec:related}
In~\cite{Maleki-Jafar} and~\cite{TDMA-TIM}, the problem of characterizing network topologies for which TDMA is optimal, was studied with no CSIT and no cooperative transmission. It was shown in~\cite{TDMA-TIM} that TDMA can be used to achieve the \emph{all-unicast} DoF region if and only if the bipartite network topology graph is chordal, i.e., every cycle that can contain a chord has one. The all-unicast setup refers to the case when each transmitter has an independent message for each receiver. This implies that if the network is chordal, then TDMA can be used to achieve the DoF region (and hence, the sum DoF) for any subset of unicast messages as well. As we will discuss in Section~\ref{sec:general}, locally connected networks are chordal, and hence, TDMA is optimal in our considered setting if cooperative transmission is not allowed, even if each message can be available at an arbitrary single transmitter.  

While the focus of this work is on studying the setting where no transmitter has access to any information about the channel state, there has been work in the literature that studies settings that are in between our setting and the traditionally considered setting of perfect CSIT. For example, in~\cite{Gesbert-2}, the authors considered a distributed CSI scenario where each transmitter has its own estimate of the channel state. In~\cite{lapidoth-conjecture}-\cite{Davoodi-Jafar}, the case where transmitters have access to a finite precision CSI was studied. 
It is also worth mentioning that even though we are assuming no CSIT in this work, we are still allowing receivers to have perfect knowledge of the channel state. In~\cite{Jafar-Goldsmith-2005}, the DoF was studied when there is no channel state information available at the receivers (no CSIR).

Finally, as part of the studied system model, we consider a constraint that allows each receiver's message to be available at few transmitters, without restriction on the identity of these transmitters. In cellular downlink, this reflects a scenario where there is a cloud controller that can assign the messages, or \emph{cell associations}, through a rate-limited backhaul. In light of recent and upcoming advances in wireless networks in general, and cellular networks in particular, this setting falls under the umbrella of Cloud Radio Access Networks (C-RAN). For examples on recent works in the literature on C-RAN systems, see~\cite{CRAN}-\cite{CRAN-Simeone-2}.

\section{System Model}\label{sec:model}
We use the standard model for the $K-$user interference channel with a single antenna at each node.
\begin{equation}\label{eq:signal}
Y_i(t) = \sum_{j=1}^{K} H_{i,j}(t) X_j(t) + Z_i(t),i\in[K],
\end{equation}
where $t$ is the time index, $X_i(t)$ is the transmitted signal of transmitter $i$, $Y_i(t)$ is the received signal of receiver $i$, $Z_i(t)$ is the zero mean unit variance Gaussian noise at receiver $i$, $H_{i,j} (t)$ is the channel coefficient from transmitter $j$ to receiver $i$ over the $t^{th}$ time slot, and $[K]$ denotes the set $\{1,2,\ldots,K\}$. In the rest of the paper, unless explicitly stated otherwise, we remove the time index for brevity.

For any set ${\cal A} \subseteq [K]$, we define the complement set $\bar{\cal A} = \{i: i\in[K], i\notin {\cal A}\}$. For each $i \in [K]$, let $W_i$ be the message (word) intended for receiver $i$. We use the abbreviations $X_{\cal A}$ and $Y_{\cal A}$ to denote the sets $\{X_i, i\in {\cal A}\}$ and $\{Y_i, i\in {\cal A}\}$, respectively.
For the $i^{\text{th}}$ user, message $W_i$ is assigned to transmitters with indices in the transmit set ${\cal T}_i \subseteq [K]$

\subsection{Channel Model}\label{sec:model}
Each transmitter is connected to its corresponding receiver as well as $L$ following receivers, and each of the last $L$ transmitters is connected to its corresponding receiver as well as all following receivers. More precisely,

\begin{equation}\label{eq:channel}
H_{i,j} \neq 0 \text { if and only if } i \in \{j,j+1,\cdots,j+L\},\forall i,j \in [K].
\end{equation}

Each non-zero channel coefficient is drawn independently from the same continuous distribution. Unless stated otherwise, all results in the paper are valid regardless of the coherence time of the channel (whether the channel remains constant across time slots or changes). While all receivers are assumed to be aware of the channel state information, the knowledge available for the design of the transmission scheme is that of the network topology. In other words, no channel state information is available at the transmitters (no CSIT). 
\subsection{Linear Cooperation Schemes}
In Section~\ref{sec:coop} of this work, we allow each message to be available at multiple transmitters, and restrict our attention to linear cooperation schemes, where the transmit signal at each transmitter is given by a linear combination of signals; each depending only on one message. More precisely,
\begin{equation}
X_j = \sum_{i: j\in{\cal T}_i} X_{j,i}, \forall j\in[K],
\end{equation}
where $X_{j,i}$ depends only on message $W_i$. 

Each message $W_i$ is represented by a vector $\mathbf{w}_i\in\mathbb{C}^{m_i}$ of $m_i$ complex symbols that are desired to be delivered to the $i^{\text{th}}$ receiver. This message is encoded to one or many of the transmit vectors $\mathbf{X}_{j,i}^n=V_{j,i}^n \mathbf{w}_i$, where $j \in {\cal T}_i$ and $V_{j,i}^n$ denotes the $n\times m_i$ linear beamforming \emph{precoding} matrix used by transmitter $j$ to transmit $W_i$ over $n$ time slots. The rank of $V_{j,i}^n$ is $m_{j,i}$, where $m_{j,i} \leq m_i$. Under such a scheme, the received signal of receiver $i$ over the $n$ time slots in (\ref{eq:signal}) can be rewritten as
\begin{eqnarray}\label{eq:rx}
\mathbf{Y}_i^n&=& \sum_{j\in\left(\{i,i-1,\cdots,i-L\} \cap {\cal T}_i\right)} H_{i,j}^n V_{j,i}^n \mathbf{w}_i\nonumber\\&+&\sum_{(k,l): k\in \left([K] \backslash \{i\}\right), l \in \{i,i-1,\cdots,i-L\} \cap {\cal T}_k} H_{i,l}^n V_{l,k}^n \mathbf{w}_k+\mathbf{Z}_i^n\nonumber\\,
\end{eqnarray}
where $\mathbf{Z}_i^n$ represents the $n \times 1$ vector of Gaussian noise values at receiver $i$ during the $n$ time slots. For every $i,j\in\{1,...,K\}$, $H_{i,j}^n$ is an $n\times n$ diagonal matrix with the $k^{\textrm{th}}$ diagonal element being equal to the value of the channel coefficient between transmitter $i$ and receiver $j$ in time slot $k$ . Each precoding matrix $V_{k,i}^n$ can only depend on the knowledge of topology. 

\subsection{Degrees of Freedom Gain in Large Networks}
The total power constraint across all the users is $P$.  The rates $R_i(P) = \frac{\log|W_i|}{n}$ are achievable if the error probabilities of all messages can be simultaneously made arbitrarily small for a large enough block length $n$. The capacity region $\mathcal{C}(P)$ is the set of all achievable rate tuples. The DoF ($\eta$) is defined as $\limsup_{P \rightarrow \infty}\frac{ C_{\Sigma}(P)}{\log P}$, where $C_\Sigma(P)$ is the sum capacity. Since $\eta$ depends on the specific choice of transmit sets, we define $\eta(K,L,\{{\cal T}_i, i\in[K]\})$ as the DoF for an $L$-connected $K$-user channel satisfying~\eqref{eq:channel}, and message $W_i$ is available at all the transmitters in ${\cal T}_i$ for each user $i\in[K]$.  We define the asymptotic per user DoF $\tau(L,\{{\cal T}_i, i\in[K]\})$ to measure how the sum degrees of freedom scales with the number of users.
\begin{equation}
\tau(L,\{{\cal T}_i, i\in[K]\}) = \lim_{K\rightarrow \infty} \frac{\eta(K,L,\{{\cal T}_i, i\in[K]\})}{K}.
\end{equation}

When each message is allowed to be available at all transmitters, we replace $\tau(L,\{{\cal T}_i=[K], \forall i\in[K]\})$ by $\tau_c(L)$ for brevity. When we impose a constraint on the maximum transmit set size, $|{\cal T}_i| \leq M, \forall i\in[K]$, and pick the choice of transmit sets that maximize the sum DoF, then we denote the asymptotic per user DoF by $\tau(L,M)$. We also add the TDMA keyword as a superscript whenever we restrict the choice of coding scheme to a TDMA scheme. For example, $\tau^{\text{TDMA}} (L=1,M=1)$ denotes the asymptotic per user DoF for the case when each transmitter is connected to the receiver with the same index as well as one following receiver, and each message can be available at any single transmitter, and only TDMA schemes can be used. It is worth noting here that modifying the channel model such that the channel coefficient $H_{i,j}$ between transmitter $j \in \{K,K-1,\cdots,K-L\}$ and receiver $i \in \{1,2,\cdots,L-(K-j)+1\}$ is non-zero (cyclic model) does not change the value of the asymptotic per user DoF.

\section{Wyner's Symmetric Network with Cooperative Transmission}\label{sec:coop}
In~\cite{ElGamal-spawc}, it was shown that linear cooperative transmission cannot increase the asymptotic per user DoF for Wyner's asymmetric model ($L=1$ case). Further, the optimal value can be achieved through a TDMA scheme. For our system model, the result of~\cite{ElGamal-spawc} is the statement that,
\begin{equation}
\tau_c(L=1) = \tau^{\text{TDMA}}(L=1,M=1)
\end{equation}
In the following theorem, we extend that conclusion to the case where $L=2$. 
\begin{thm}\label{thm:main}
For a locally connected channel with connectivity parameter $L=2$, if each message can be available at all transmitters and linear cooperative transmission is allowed, the optimal value of the asymptotic per user DoF is achieved by assigning each message to one transmitter and using a TDMA scheme.
\begin{equation}
\tau_c(L=2) = \tau^{\text{TDMA}}(L=2,M=1) = \frac{1}{2}
\end{equation}
\end{thm}
We spend the rest of this section proving Theorem~\ref{thm:main}. Since TDMA schemes are linear schemes, it is straightforward to see that $\tau_c(L=2) \geq \tau^{\text{TDMA}}(L=2,M=1)$. Also, we can see that $\tau^{\text{TDMA}}(L=2,M=1) \geq \frac{1}{2}$ by using the message assignment ${\cal T}_i = \{i-1\}$ for each $i \in [K]$, and turning off every other transmitter. Hence, it suffices to prove that $\tau_c(L=2) \leq \frac{1}{2}$.

We show that the DoF of any network with an even number of users $K$, is at most $\frac{K}{2}$, even if each message is available at all transmitters. We use the following result in \cite[Lemma $4$]{ElGamal-Annapureddy-Veeravalli-IT14}.

\begin{lem}\label{lem:dofouterbound}
If there exists a set ${\cal B}\subseteq [K]$, a function $f_1$, and a function $f_2$ whose definition does not depend on the transmit power constraint $P$, and $f_1\left(Y_{\cal B},X_{U_{\cal B}}\right)=X_{\bar{U}_{\cal B}}+f_2(Z_{\cal B})$, then the sum DoF $\eta \leq |{\cal B}|$. 
\end{lem} 
In the above lemma, we used $U_{\cal B}$ as the set of indices of transmitters that exclusively carry the messages for the receivers in ${\cal B}$, and the complement set $\bar{U}_{\cal B}$ as the set of indices of transmitters that carry messages for receivers outside ${\cal B}$. More precisely, $U_{\cal B} = [K]\backslash\cup_{i \notin {\cal B}} {\cal T}_i$. 

In order to prove Theorem~\ref{thm:main}, we use Lemma~\ref{lem:dofouterbound} with the set ${\cal B}=\{i \in [K]: i \text{ mod } 2=0\}$. In other words, ${\cal B}$ is the set of even indices, $Y_{\cal B}$ is the set of received signals with even indices, and $X_{U_{\cal B}}=\{X_{j,i}: i\in{\cal B}, j\in{\cal T}_i\}$ is the set of transmit signals that depend on messages with even indices. Let $f_2(Z_{\cal B})=-Z_{\cal B}$, then it suffices to show how to reconstruct the set of transmit signals $X_{\bar{U}_{\cal B}}=\{X_{j,i}: i\in\bar{\cal B}, j\in{\cal T}_i\}$ from the processed received signals $\tilde{Y}_{\cal B}$ that are obtained by removing the contributions of $Z_{\cal B}$ and $X_{U_{\cal B}}$ from $Y_{B}$, where for each $i\in {\cal B}$ the following holds.

\begin{equation}\label{eq:one}
{\bf \tilde{Y}}_i^n = {\bf Y_i}^n - {\bf Z}^n_i - \sum_{k \in {\cal B}, l \in {\cal T}_k \cap \{i,i-1,i-2\}} H_{i,l}^n V_{l,k}^n {\bf w}_k
\end{equation}
We notice that $\tilde{Y}_{\cal B}$ is a linear combination of the transmit signals $X_{\bar{U}_{\cal B}}$. What remains to show is that this linear transformation is full rank. More precisely, we obtain the following for each $i \in {\cal B}$ from~\eqref{eq:rx} and~\eqref{eq:one}.
\begin{equation}\label{eq:two}
{\bf \tilde{Y}}_i^n=\sum_{k \in \bar{\cal B}, l \in {\cal T}_k \cap \{i,i-1,i-2\}} H_{i,l}^n V_{l,k}^n {\bf w}_k.
\end{equation}
If we assume that $X_{j,i}=0$ for any message $i \in [K]$ and a transmitter $j \in [K]: j \notin {\cal T}_i$ that is not in the transmit set the message, then $~\eqref{eq:two}$ can be rewritten as,
\begin{equation}
{\bf \tilde{Y}}_i^n=\sum_{k \in \bar{\cal B}, l \in \{i,i-1,i-2\}} H_{i,l}^n V_{l,k}^n {\bf w}_k.
\end{equation}
In matrix form, the signals ${\bf \tilde{Y}}_{\cal B}^n$ can be written as,
\begin{equation}\label{eq:yone}
{\bf \tilde{Y}}_{\cal B}^n = H_{{\cal B},[K]}^n V_{[K],\bar{\cal B}}^n {\bf w}_{\bar{\cal B}},
\end{equation}
where ${\bf w}_{\bar{\cal B}}$ is an $\frac{nK}{2} \times 1$ vector containing the symbols corresponding to the $\frac{K}{2}$ messages with indices in ${\bar{\cal B}}$ over $n$ time slots. Note that each message will have at most $n$ symbols over $n$ time slots. Also, $V_{[K],\bar{\cal B}}^n$ is an $nK \times \frac{nK}{2}$ matrix containing the transmit beamforming coefficients used by each of the $K$ transmitters to each generate a vector of $n$ transmit signals over $n$ time slots, that are used to transmit the $\frac{nK}{2}$ symbols ${\bf w}_{\bar{\cal B}}$. Finally, $H_{{\cal B},[K]}^n$ is an $\frac{nK}{2} \times nK$ matrix that contains the channel coefficients between  each of the $\frac{K}{2}$ receivers in ${\cal B}$ and each of the $K$ transmitters over $n$ time slots. The matrix $H_{{\cal B},[K]}^n$ has the form illustrated in~\eqref{eq:matrix_one} below, where we use the notation $H_{i,j}^{(k)}$ to denote the channel coefficient between transmitter $j$ and receiver $i$ over time slot $k$. 

Similarly, the processed received signals for the complement set ${\bf \tilde{Y}}_{\bar{\cal B}}^n$ are given by,
\begin{equation}\label{eq:ytwo}
{\bf \tilde{Y}}_{\bar{\cal B}}^n = H_{\bar{\cal B},[K]}^n V_{[K],\bar{\cal B}}^n {\bf w}_{\bar{\cal B}},
\end{equation}
where $H_{\bar{\cal B},[K]}^n$ is given in~\eqref{eq:matrix_two} below.

\begin{figure*}[!t]
\normalsize
\setcounter{MaxMatrixCols}{20}
\begin{equation}
\label{eq:matrix_one}
H_{{\cal B},[K]}^n = 
\begin{bmatrix}
H_{2,1}^{(1)} & \hdots & H_{2,1}^{(n)} & H_{2,2}^{(1)} & \hdots& H_{2,2}^{(n)} & 0 &\hdots & \hdots & \hdots& \hdots & \hdots & \hdots &\hdots &\hdots & 0 \\
0 & \hdots  & 0 & H_{4,2}^{(1)} & \hdots & H_{4,2}^{(n)} & H_{4,3}^{(1)} & \hdots & H_{4,3}^{(n)}& H_{4,4}^{(1)} & \hdots & H_{4,4}^{(n)} & 0 &\hdots & \hdots & 0 \\
\vdots \\
0 & \hdots & \hdots&\hdots & \hdots&\hdots&\hdots&\hdots&\hdots&\hdots&\hdots&\hdots&\hdots& H_{K,K-2}^{(1)} & \hdots & H_{K,K}^{(n)}\\
\end{bmatrix},
\end{equation}

\begin{equation}
\label{eq:matrix_two}
H_{\bar{\cal B},[K]}^n = 
\begin{bmatrix}
H_{1,1}^{(1)} & \hdots & H_{1,1}^{(n)} & 0 &\hdots & \hdots & \hdots& \hdots & \hdots & \hdots & \hdots & \hdots & \hdots &\hdots &\hdots & 0 \\
H_{3,1}^{(1)} & \hdots & H_{3,1}^{(n)} & H_{3,2}^{(1)} & \hdots & H_{3,2}^{(n)}& H_{3,3}^{(1)} & \hdots & H_{3,3}^{(n)} & 0 &\hdots & \hdots & \hdots & \hdots & \hdots & 0 \\
\vdots \\
0 & \hdots&\hdots&\hdots&\hdots&\hdots&\hdots&\hdots&\hdots&\hdots& H_{K-1,K-3}^{(1)} & \hdots & H_{K-1,K-1}^{(n)}&0&\hdots&0\\
\end{bmatrix},
\end{equation}
\hrulefill
\vspace*{4pt}
\end{figure*}

Note that our converse proof must be oblivious to the choice of transmit beamforming matrix $V_{[K],\bar{\cal B}}^n$, and all we know about it is that it cannot depend on the channel coefficients. We finish the proof by showing that one of two scenarios has to occur. The first scenario is when ${\bf w}_{\bar{\cal B}}$ can be recovered from ${\bf \tilde{Y}}_{\cal B}^n$ through~\eqref{eq:yone}, i.e., the rank of $H_{{\cal B},[K]}^n V_{[K],\bar{\cal B}}^n$ equals the number of non-zero entries in ${\bf w}_{\bar{\cal B}}$ for almost all realizations of the channel coefficients. The second scenario is when $H_{{\cal B},[K]}^n V_{[K],\bar{\cal B}}^n$ has rank deficiency. In this case, we show that we can obtain a number of rows in the matrix $H_{\bar{\cal B},[K]}^n V_{[K],\bar{\cal B}}^n$ from $H_{{\cal B},[K]}^n V_{[K],\bar{\cal B}}^n$, and that number is equal to the rank deficiency of $H_{{\cal B},[K]}^n V_{[K],\bar{\cal B}}^n$. In other words, if we let $s$ be the number of non-zero entries in ${\bf w}_{\bar{\cal B}}$, and assume that ${\bf \tilde{Y}}_{\cal B}^n$ has only a number $r$ of linearly independent equations in ${\bf w}_{\bar{\cal B}}$, then in this case we show that we can obtain \emph{statistically equivalent} versions of $s-r$ signals of  ${\bf \tilde{Y}}_{\bar{\cal B}}^n$. Further, from these $s-r$ signals, we can decode $s-r$ symbols of ${\bf w}_{\bar{\cal B}}$, and hence, the remaining symbols can be decoded from~\eqref{eq:yone}. By \emph{statistically equivalent}, we are exploiting here the fact that the transmit beamforming matrix does not depend on the channel coefficients, and hence, obtaining for example a signal $H_{2,1}X_1 + H_{2,2}X_2$ is statistically equivalent to the signal $H_{3,1}X_1+H_{3,2}X_2$. We provide the formal proof of this part in the journal version of this work.

\section{Discussion: General Locally Connected Networks}\label{sec:general}
Imposing the constraint $M=1$ is equivalent to allowing for a flexible association of messages to transmitters, while disabling cooperative transmission. It was shown in~\cite{TDMA-TIM} that TDMA can be used to achieve the no CSIT DoF region for any chordal network topology, with any arbitrary subset of unicast messages. Since all locally connected networks are chordal, the following statement holds as a corollary to the result in~\cite{TDMA-TIM},
\begin{cor}\label{cor:tdma}
If cooperative transmission is not allowed, then TDMA is optimal for all considered locally connected networks.
\begin{equation}
\tau(L,M=1)=\tau^{\text{TDMA}}(L,M=1), \forall L\in{\bf Z}^+.
\end{equation}
\end{cor}
In fact, the same conclusion of Corollary~\ref{cor:tdma} would still hold if cooperation is only allowed through splitting messages into independent parts, and distributing different parts to different transmitters. Hence, the open questions remaining are about the values $\tau_c(L)$ and $\tau(L,M)$ for $L > 2$ and $M > 1$; specifically when more than one transmitter are using common information about a message.   

In~\cite{jafar-topological}, \cite{demand-graph-1} and \cite{demand-graph-2}, a key result was found for bounding the sum DoF when each message can only be available at a single transmitter. It was shown that the sum DoF of any subset of messages that form an \emph{acyclic demand graph} is unity. A demand graph is a directed bipatite graph, where one partite set has messages and the other partite set has receivers. An edge exists from a message to a receiver if the message is destined at the receiver. An edge exists from a receiver to a message if the receiver is not connected to the transmitter carrying the message. In our setting, we already know that message $W_i$ is destined for the $i^{\text{th}}$ receiver, then we can simply collapse each message-destination pair into one node and still have the same conclusion about cycles. Further, since we know that the $i^{\text{th}}$ transmitter is connected to receivers with indices $\{i,i+1,\cdots,i+L\}$, then if we take any $L+2$ nodes with consecutive indices, there will be a cycle between the first and last nodes, and hence the sum DoF cannot be bounded by unity. This provides a simple way to explain why the following holds. 
\begin{equation}\label{eq:mone}
\tau(L,M=1)=\frac{2}{L+2}, \forall L \in {\bf Z}^+.
\end{equation}
It was shown in~\cite{ElGamal-Annapureddy-Veeravalli-IT14} that the expression in~\eqref{eq:mone} extends to $\frac{2M}{2M+L}$ when each message can be available at $M$ transmitters, and only zero-forcing transmit beamforming is allowed, and CSIT is available. We are hoping to obtain an analogous generalization with no CSIT and with the restriction to linear cooperation schemes, by identifying a key lemma for bounding the sum DoF of a subset of messages akin to the one discussed above, but with each message available at multiple transmitters.

Finally, we would like to point out to a key difficulty with extending the result we have for $L=1$ in~\cite{ElGamal-spawc} and $L=2$ in this work, to more general values of $L > 2$. When deriving a converse for a puDoF value that is greater than or equal to half, we rely on the fact that when we apply Lemma~\ref{lem:dofouterbound}, it is possible to choose the set ${\cal B}$ such that $|{\cal B}| \geq |\bar{\cal B}|$. Hence, we can have a number of given equations that is at least equal to the number of missing variables, and that enables a valid application of the lemma. Now, for $L > 2$, $\tau^{\text{TDMA}} (L,M=1) < \frac{1}{2}$. If it is true that $\tau(L>2,M>1)=\tau^{\text{TDMA}}(L,1)$, then the converse argument would involve more than a direct application of Lemma~\ref{lem:dofouterbound}.

\section{Concluding Remarks}\label{sec:conclusion}
We have studied, through the asymptotic per user DoF criterion, the question of whether linear cooperation schemes are useful in locally connected wireless networks with no channel state information available at any transmitter (no CSIT). We have shown that for the case when each receiver observes two interfering signals (Wyner's symmetric network), assigning each message to a single transmitter and using TDMA is optimal even if each message can be available at all transmitters and linear cooperative transmission is allowed. We have then highlighted that existing work in the literature imply that TDMA is optimal for a general locally connected channel, if each message can only be available at a single transmitter. 

The answer to our question remains open in a general setting. It is not clear whether the obtained results are particular to Wyner's asymmetric and symmetric models of channel connectivity, or if it is true in general that TDMA is optimal, as long as the transmitters are not aware of the channel state information, even if cooperative transmission is allowed?

\section*{}

\end{document}